\documentclass[%
reprint,
 amsmath,amssymb,
 aps, showkeys,
]{revtex4-2}

\usepackage{amsmath,amssymb,amsfonts}
\usepackage{algorithmic}
\usepackage{graphicx}
\usepackage{gensymb}
\usepackage{textcomp}
\usepackage{multirow}
\usepackage{balance}
\usepackage{physics}
\usepackage[table]{xcolor}
\def\BibTeX{{\rm B\kern-.05em{\sc i\kern-.025em b}\kern-.08em
    T\kern-.1667em\lower.7ex\hbox{E}\kern-.125emX}}

\begin{document}

\preprint{APS/123-QED}

\title{Integrated Differential Conjugate Homodyne Detection for Quantum Random Number Generation}
\thanks{Funding for this work was provided by the U.S. Department of Energy, Office of Cybersecurity Energy Security and Emergency Response (CESER) through the Risk Management Tools and Technologies (RMT) Program. This manuscript has been supported by UT-Battelle, LLC under Contract No. DEAC05-00OR22725 with the U.S. Department of Energy. This work was supported in part by onsemi for chip fabrication.}

\author{Christian Carver, Jared Marchant, Benjamin Fisher, Nicholas Townsend, Tyler Stowell, Austin Barlow, Benjamin Arnesen, Shiuh-Hua Wood Chiang, Ryan M. Camacho}
\affiliation{Department of Electrical and Computer Engineering, Brigham Young University, Provo, UT 84602 USA}

\date{December 2, 2024}

\begin{abstract}
In this work, we perform on-chip quantum random number generation (QRNG) that uses a novel differential amplifier configuration for conjugate homodyne detection. Leveraging separate integrated photonics and integrated analog circuit platforms, we present an alternative method for QRNG. This approach exploits the observable $\hat{\text{Z}}$, derived from the sum of squared conjugate quadrature distributions which we compare to the traditional single quadrature approach. Utilizing this method, we report a shot noise clearance (SNC) of 25.6 dB and a common mode rejection ratio (CMRR) of 69 dB for our homodyne detection system. We used a variety of design tools to model and predict performance and compare results with our measurements. The realization of our QRNG system consists of a 90° optical hybrid, a dual differential transimpedance amplifier (TIA), and a field-programmable gate array (FPGA) used for the real-time post-processing to produce a uniform random bitstream. The randomness extraction is implemented using a Toeplitz hashing algorithm and is validated by the National Institute of Standards and Technology (NIST) randomness test suites.
\end{abstract}

\keywords{analog-to-digital converter (ADC), analog integrated circuit (AIC), differential transimpedance amplifier (TIA), electronic integrated circuit (EIC), field-programmable gate array (FPGA), photonic integrated circuit (PIC), quantum random number generation (QRNG)}

\maketitle

\section{\label{sec:introduction}Introduction}

Quantum random number generation (QRNG) is an important tool for various communication and computation primitives. To date, many of the highest-speed QRNG implementations amplify vacuum fluctuations using homodyne detection \cite{bruynsteen2023100, integrated-homodyne-best, tanizawa2024real, bai202118, haylock2019multiplexed}. This scheme is advantageous, as vacuum fluctuations have zero-mean amplitude and phase, resulting in phase-independent amplification. However, there are some potential disadvantages. First, the performance of this scheme is very sensitive to splitting ratio imbalance in homodyne measurements. Second, the probability density function of the resulting random noise is limited to that of a vacuum state. This may be undesirable when maximizing the minimum entropy for QRNG or maximizing the signal-to-noise ratio (SNR) for more advanced quantum networking functions beyond QRNG, such as quantum key distribution (QKD) or quantum state tomography.

In this paper, we propose a new scheme for QRNG based on conjugate homodyne detection and a novel differential amplification approach for homodyne detection. This scheme uses an active power balancing technique to ensure an optimal splitting ratio. Additionally, the conjugate homodyne detection scheme introduced by Bing, et al. \cite{bing} allows for a phase independent measurement of states other than the vacuum. We build upon that work by integrating the system onto a photonic chip platform and provide an example application for QRNG.

Our approach produces an SNR comparable to previous state-of-the-art work, allowing for a high theoretical QRNG bitrate. Unlike previous research, we measure the $\hat{\text{Z}}$ quantum operator, which relates to our homodyne measurements via $\hat{\text{Z}} = \hat{\text{X}}^2 + \hat{\text{P}}^2$. This observable’s distribution is tied to the photon number distribution and has the clear advantage of being phase independent, facilitating easier experimental measurements. We conducted experiments with a custom dual-chip setup: a photonic integrated circuit (PIC) with on-chip photodiodes and a differential transimpedance amplifier (TIA).

\begin{figure*}[t!]
\includegraphics[width=\linewidth]{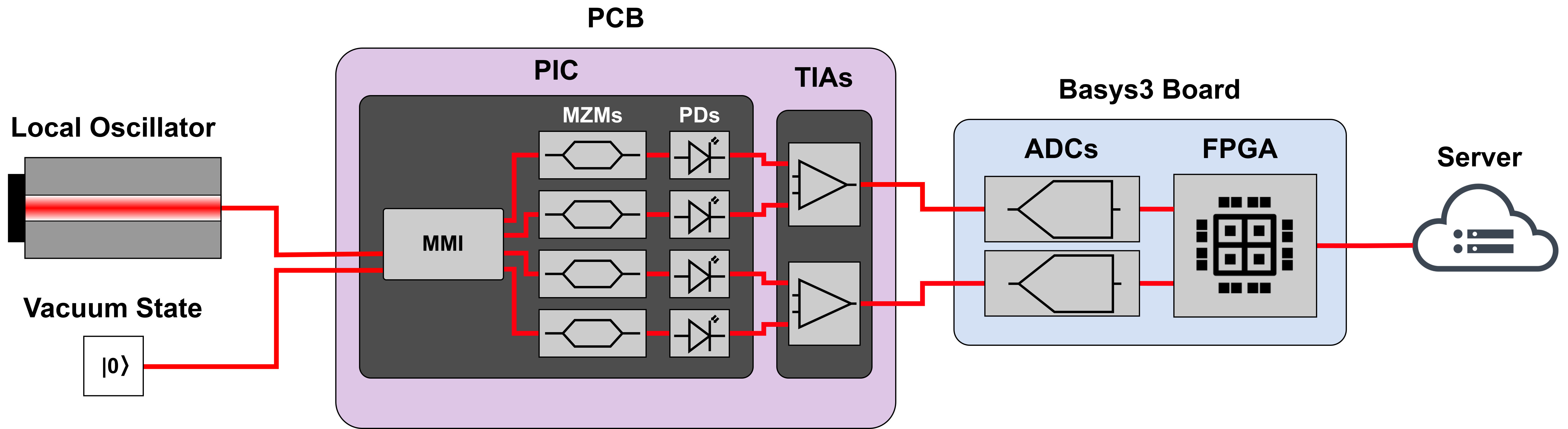}
\caption{\label{fig:block_diagram}Block Diagram of Quantum Random Number Generation (QRNG) System. The physical system consists of a printed circuit board (PCB), which integrates a photonic integrated circuit (PIC) and transimpedance amplifiers (TIAs) via wirebonding, and a Basys3 development board, which hosts two analog-to-digital converters (ADCs) and a field-programmable gate array (FPGA). Components contained on the PIC include a multi-mode interferometer (MMI), Mach–Zehnder modulators (MZMs), and germanium photodiodes (PDs).}
\end{figure*}

\section{System Overview}
The PIC and TIA form the basis of our QRNG system, as shown in Fig. \ref{fig:block_diagram}. On the PIC, a strong local oscillator (LO) is mixed with a quantum signal using a 90° optical hybrid. The PIC interferes these input signals and performs conjugate homodyne detection, which produces Gaussian distributions for the $\hat{\text{X}}$ and $\hat{\text{P}}$ quadratures. These signals are amplified by the differential TIA and sent to a field-programmable gate array (FPGA) via an analog-to-digital converter (ADC). The FPGA implements a Toeplitz hashing extractor to produce a uniform random number distribution from only the quantum entropy in $\hat{\text{X}}$ and $\hat{\text{P}}$ \cite{ma2013postprocessing}, quantities that are traditionally used in QRNG systems. We compare this to our novel approach using the $\hat{\text{Z}}$ distribution. The random numbers are then available via a public API to request quantum random numbers in real time from our laboratory. These numbers can be requested anytime at camacholab.byu.edu/qrng.

\section{Design Tools}
\subsection{Quantum PIC Simulation}
When developing a PIC, designers use a variety of tools to accurately simulate device performance before fabrication. 3-D finite-difference time-domain (FDTD) solvers are optimal for predicting device performance but do not scale very well with large circuits, as they have a time complexity of $O(n^4)$, where $n$ relates to the simulation volume. For this reason, designers often only use FDTD to optimize single components. Designers can obtain S-parameters for individual components from these simulations to then use them in a more optimized circuit solver to predict the circuit performance. 

Most photonic circuit solvers have only classical simulation mechanics. Because our circuit relies on quantum principles, we developed our own photonic circuit solver that realizes classical S-parameters as quantum compatible unitary transformations \cite{carver2022simphony}. These S-parameters can be generated using classical FDTD simulations or measured from physical devices. We used this to optimize our homodyne detection performance and compare our final results with the simulated values.

\subsection{Electronic TIA Simulation}
We also performed electronic circuit simulations to verify TIA performance. In addition, we used Verilog A models to characterize the interactions of some of the electro-optic PIC components like the photodiodes and Mach-Zehnder modulators (MZMs) to validate that the entire system would work together as intended.

\section{Photonic Integrated Circuit}  
\subsection{Design}

\begin{figure}[b]
\includegraphics[width=0.45\textwidth]{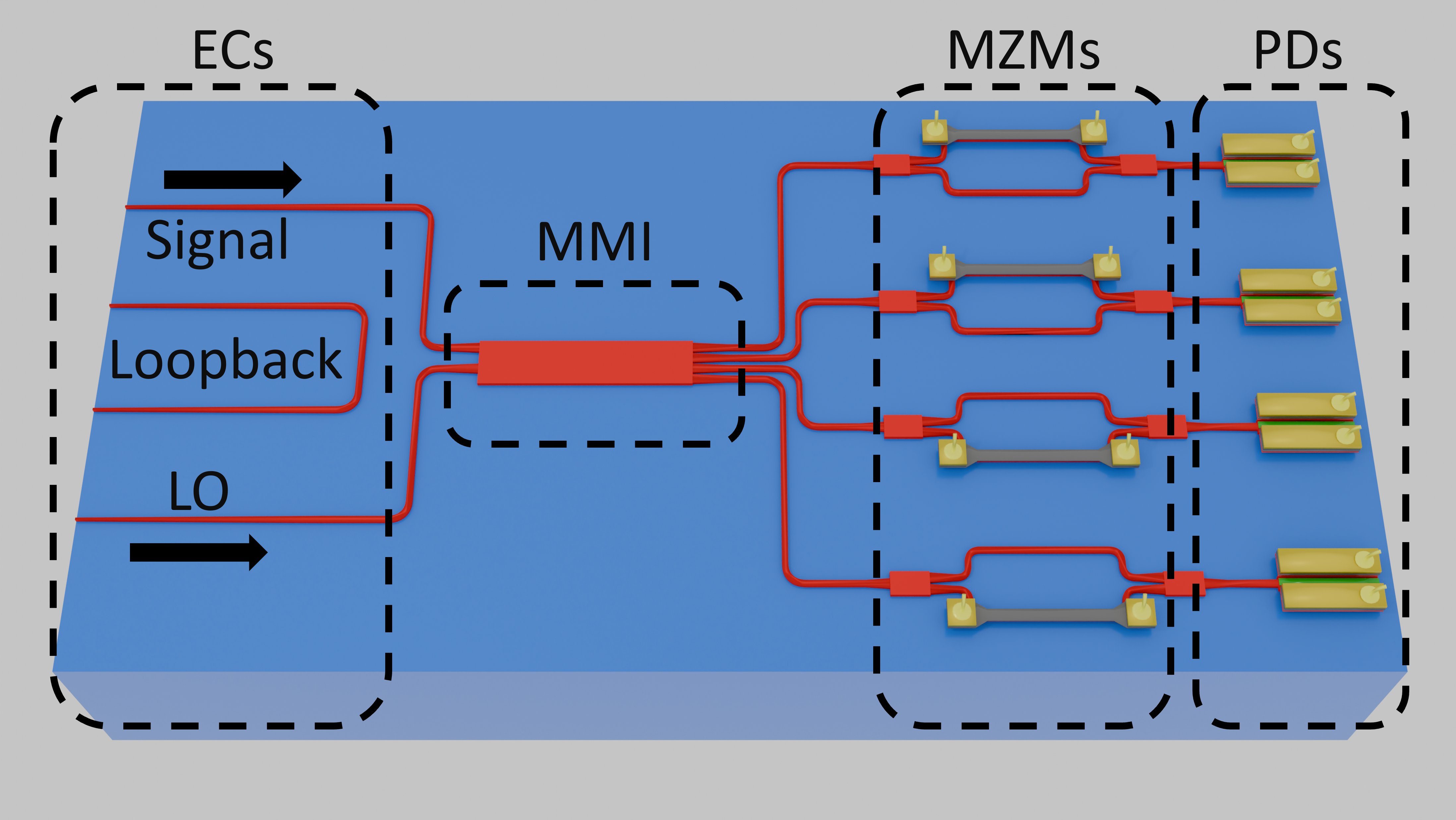}
\caption{\label{fig:gds} Schematic of Photonic Integrated Circuit (PIC). Edge couplers (ECs) guide a quantum signal and a reference onto the PIC, where they are combined in a 90$\degree$ optical hybrid, physically realized as a multi-mode interferometer (MMI). Resulting signal powers are balanced using Mach-Zehnder modulators (MZMs) and converted to electrical signals via germanium photodiodes (PDs). A loopback structure is included for fiber-chip alignment.}
\end{figure}

\begin{figure*}[t!]
\includegraphics[width=\linewidth]{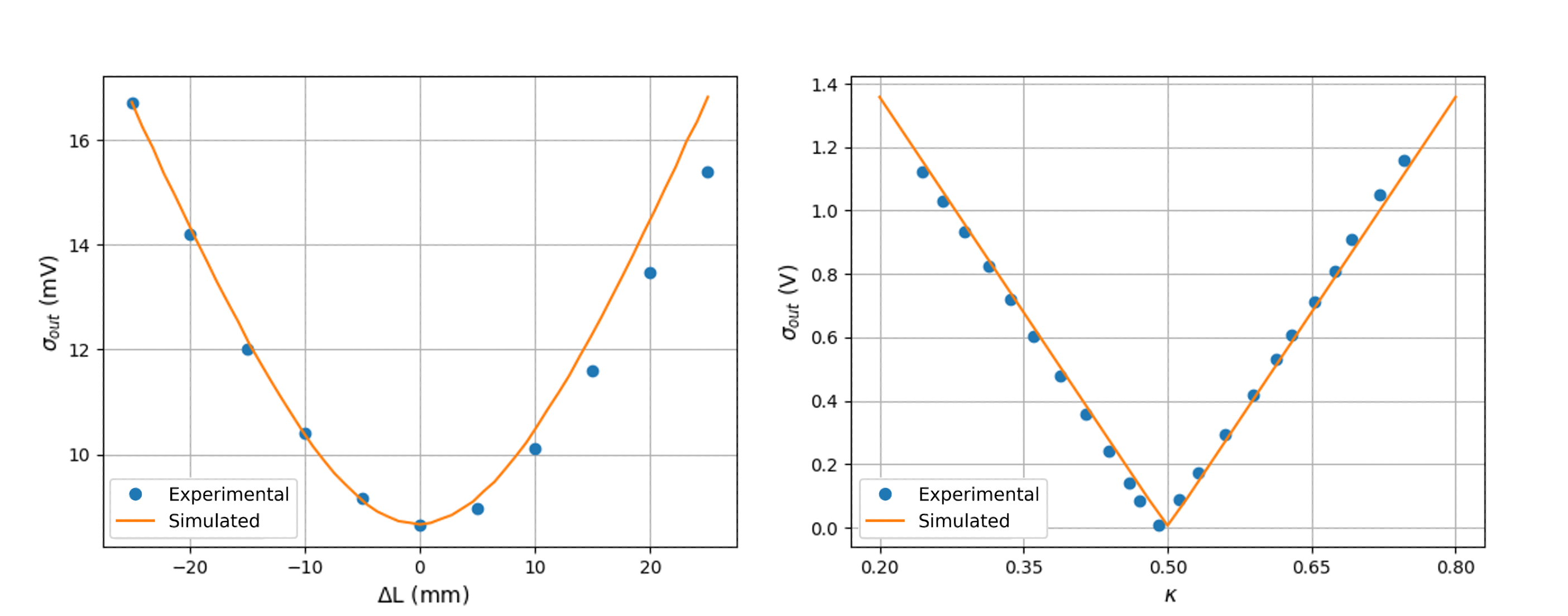}
\caption{\label{fig:pathlen}Effect of Optical Path Length on Signal Variance (left). Effect of Splitter Ratio Imbalance on Signal  Variance (right).}
\end{figure*}

A common physical realization of the 90° optical hybrid used to combine an LO and a quantum signal in a QRNG system employs a series of beamsplitters in free space optics. However, with integrated photonics this can instead be realized with either a circuit consisting of directional couplers and a y-branch or a single multi-mode interferometer (MMI), which is the method used here. A layout of the PIC is shown in Fig. \ref{fig:gds}, implemented using a silicon-on-insulator platform. Realizing QRNG using integrated circuitry not only allows for high speed designs, but also shows that QRNG can be implemented using a minimal device footprint. The four outputs of the optical hybrid, denoted X$_{1}$, X$_{2}$, P$_{1}$, and P$_{2}$ are equally spaced on the circumference of the complex unit circle. Differential transimpedance amplification is performed on the difference between the photocurrents of X$_{1}$ and X$_{2}$, as well as P$_{1}$ and P$_{2}$ (see Fig. \ref{fig:PD_subtraction}). The resulting amplified voltages correspond to the quadratures $\hat{\text{X}}$ and $\hat{\text{P}}$, respectively. The measurements of $\hat{\text{X}}$ and $\hat{\text{P}}$ are then squared and summed to obtain a measurement of $\hat{\text{Z}}$.

\subsection{Optical Path Length}
The optical common mode noise from the LO in our $\hat{\text{X}}$ and $\hat{\text{P}}$ measurements can be reduced with a large optical common mode rejection ratio (CMRR). Ideally, the difference in equal-amplitude, in-phase photocurrents reduces common mode noise to zero. One motivation for designing the system on chip is that the nanoscale precision of integrated photonic devices permits a much higher accuracy in path length balancing compared to free space devices. 

The optical CMRR due to path length difference, $\Delta L$, can be expressed as
\begin{equation}
    \text{CMRR} = -10\log_{10}\left( \abs{\sin\left( \frac{ \Delta \phi }{2}\right)} \right),
\end{equation}
where $\Delta \phi$ is the phase incurred from the path length difference, defined as
\begin{equation}
    \Delta \phi =\frac{ 2\pi f_{cm}}{c} n_{eff} \Delta L.
\end{equation}
Here, $f_{cm}$ represents the common mode frequency, $n_{eff}$ is the effective index of the guided light, and $c$ is the speed of light. See Appendix \ref{sec:appendix_a} for a complete derivation of this equation. To quantify this experimentally, we used fiber optic components to conduct a simple vacuum state homodyne detection experiment using an amplitude modulated LO, a variable optical delay line on one of the output paths, and a balance detector. We varied $\Delta L$ from -25 to +25 mm in free space. A negative $\Delta L$ corresponds to a smaller free space gap in the optical delay line, while a positive $\Delta L$ corresponds to a larger free space gap. The effect of path length difference on $\sigma_{out}$, the standard deviation of the TIA output signal, is shown in Fig. \ref{fig:pathlen} (left). The best optical CMRR is achieved when $\Delta L$ is close to zero, which limits the standard deviation to the detector's noise floor at 8.5 mV.

We compared these results to a simulation using an LO power of 33 mW and modulation frequency of 1 GHz. The experimental results deviate from simulation at greater path length differences, which can be attributed to excess loss in the variable delay line. 

\subsection{Optical Splitting Ratio and Photodiode Responsivity}
Optical CMRR is also affected by unbalanced optical splitting ratios and photodiode responsivities. In practice, high CMRR is difficult to achieve without a form of active tuning because of imperfect tolerances in the fabrication process for both splitters and photodiodes. In our design, we use an MZM before each photodiode to ensure balanced photocurrents.

The optical CMRR due to optical splitting imbalances can be expressed as
\begin{equation}
    \text{CMRR} = -10\log_{10}(|2\kappa-1|),
\end{equation} 
where $\kappa$ is the power coupling coefficient which includes the optical splitting and differences in photodiode responsivity. See Appendix \ref{sec:appendix_a} for a complete derivation of this equation. To quantify this experimentally, we modified the previous experiment by instead varying $\kappa$ from 0.25 to 0.75 while keeping the total power measured by the photodiodes constant. Fig. \ref{fig:pathlen} (right) shows that $\sigma_{out}$ is minimized when $\kappa=0.5$, or when the powers on each arm are approximately equal. We compare these results to a simulation with a power of 450 $\mu$W and modulation frequency of 500 kHz. The simulation shows good agreement with experimental values.

\begin{figure}[t!]
\includegraphics[width=0.45\textwidth]{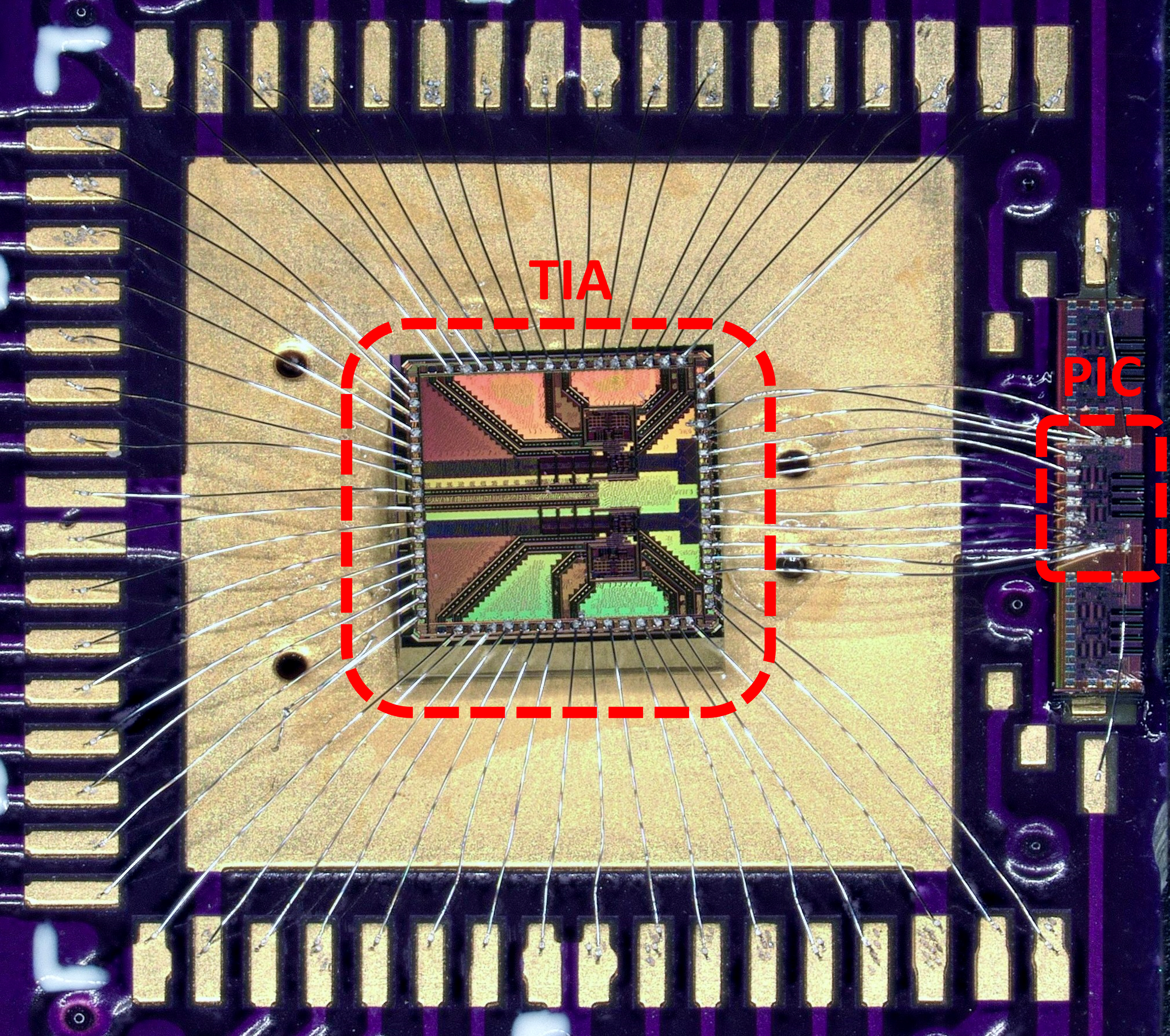}
\caption{\label{fig:wirebonds}Transimpedance Amplifier (TIA) and Photonic Integrated Circuit (PIC). The two components are interfaced on the printed circuit board (PCB) via wirebonding.}
\end{figure}

\section{Differential Transimpedance Amplifier}
After optimizing the optical components of the PIC, it is necessary to design a sufficiently low-noise transimpedance amplifier (TIA) to accurately measure the quantum signals. To accomplish this, our design uses a novel true differential amplification scheme for homodyne detection. This section explores the key design considerations for developing this amplifier, including thermal noise sources,  balancing bandwidth tradeoffs, and maximizing electrical CMRR.

Amplifier thermal noise performance is dominated by the first few transistors seen in the signal path. The noise introduced by these input metal-oxide-semiconductor field-effect transistors (MOSFETs) can be reduced by increasing their size. However, there is a direct tradeoff between bandwidth and noise performance. For resistive feedback op-amp-style TIAs, the dominant pole in the frequency response is typically at the input of the TIA. To achieve both low-noise and high-bandwidth performance we implemented additional system level techniques to reduce capacitive loading at the TIA input node.

The capacitance at the input node of the TIA typically consists of parasitic capacitance from photodiode drivers, PIC pads, printed circuit board (PCB) traces, IC packaging, IC pads, and amplifier input MOSFETs. This is usually dominated by chip pad packaging and PCB capacitance. To significantly reduce our parasitic capacitance we omitted chip packaging and PCB routing by wirebonding directly from chip to chip, shown in Fig. \ref{fig:wirebonds}. This allowed us to achieve our noise performance by re-sizing the input MOSFETs without significantly degrading bandwidth. Given these tradeoffs, the final bandwidth was designed to be 100 MHz.

\begin{figure}[t!]
\includegraphics[width=0.45\textwidth]{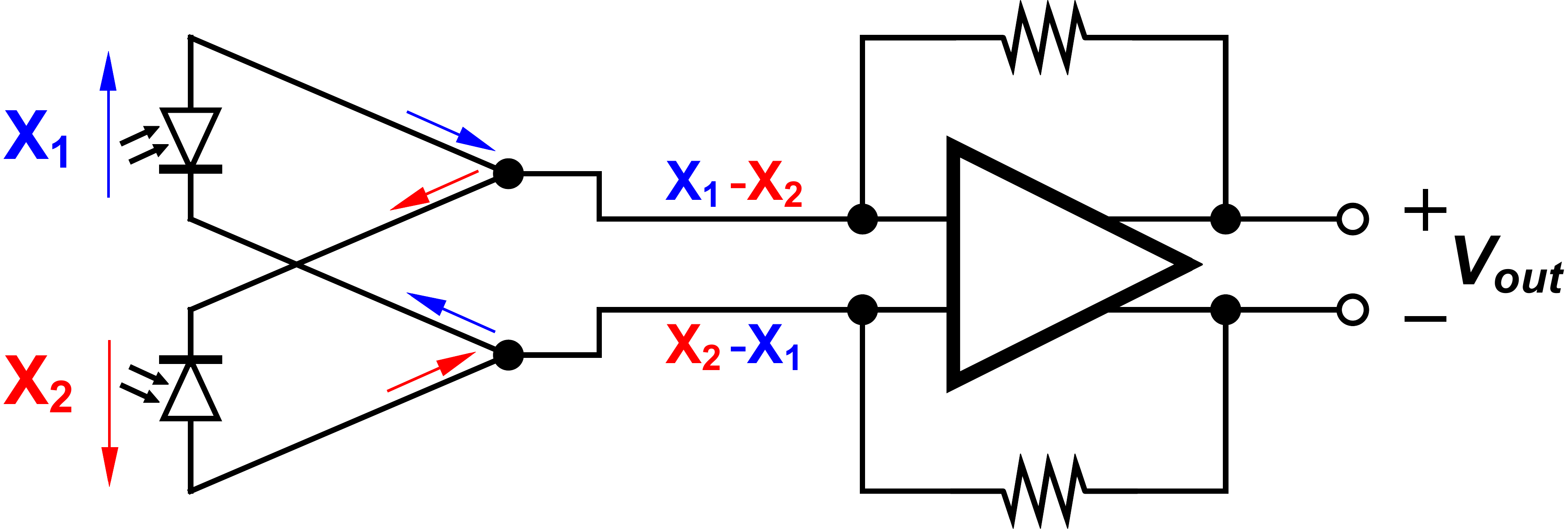}
\caption{\label{fig:PD_subtraction}Differential Transimpedance Amplification Schematic for Homodyne Detection. The photocurrents X$_1$ and X$_2$ and their corresponding current directions are shown incident to the differential amplifier, where the output $V_{out}$ is the amplified difference between X$_1$ and X$_2$.}
\end{figure}

A differential design was chosen for this architecture to improve system resilience to external noise sources. With careful design, signals that are coupled into the signal path will be coupled into each of the differential channels with approximately equal magnitude. Thus, after measurement, the two differential signals can be subtracted from each other to double the signal and remove common mode noise. By designing the TIA in conjunction with the PIC, we were able to extend the differential portion of the circuit to the photodiodes, allowing for a novel true differential design \cite{TIA}, rather than a pseudo-differential design. Capturing both positive and negative currents from each photodiode enables us to implicitly implement the necessary signal processing operations while keeping a true differential design. This is shown in Fig. \ref{fig:PD_subtraction}. Summing currents from the X$_{1}$ cathode and X$_{2}$ anode yƒields X$_1 - \text{X}_2$. Conversely, we sum currents from the X$_{1}$ anode and X$_{2}$ cathode to obtain $-(\text{X}_1 - \text{X}_2)$. We then pass these current signals into the differential TIA. After amplification and measurement we subtract the differential signals and are left with $2(\text{X}_1 - \text{X}_2)$.

\section{System Performance}
\subsection{Common Mode Rejection Ratio}
We define the CMRR of our homodyne detector by how well the detector is able to eliminate a common mode signal. One can determine the CMRR by modulating a signal on the LO and measuring the amount of that signal present after homodyne detection. CMRR is calculated using
\begin{equation}
\text{CMRR} = 10\log_{10}\left(\frac{S_{d}}{S_{cm}}\right),
\label{CMRR}
\end{equation}
where $S_{cm}$ is the output common mode signal amplitude and $S_{d}$ is the input differential signal amplitude. 

\begin{figure}[t!]
\includegraphics[width=0.45\textwidth]{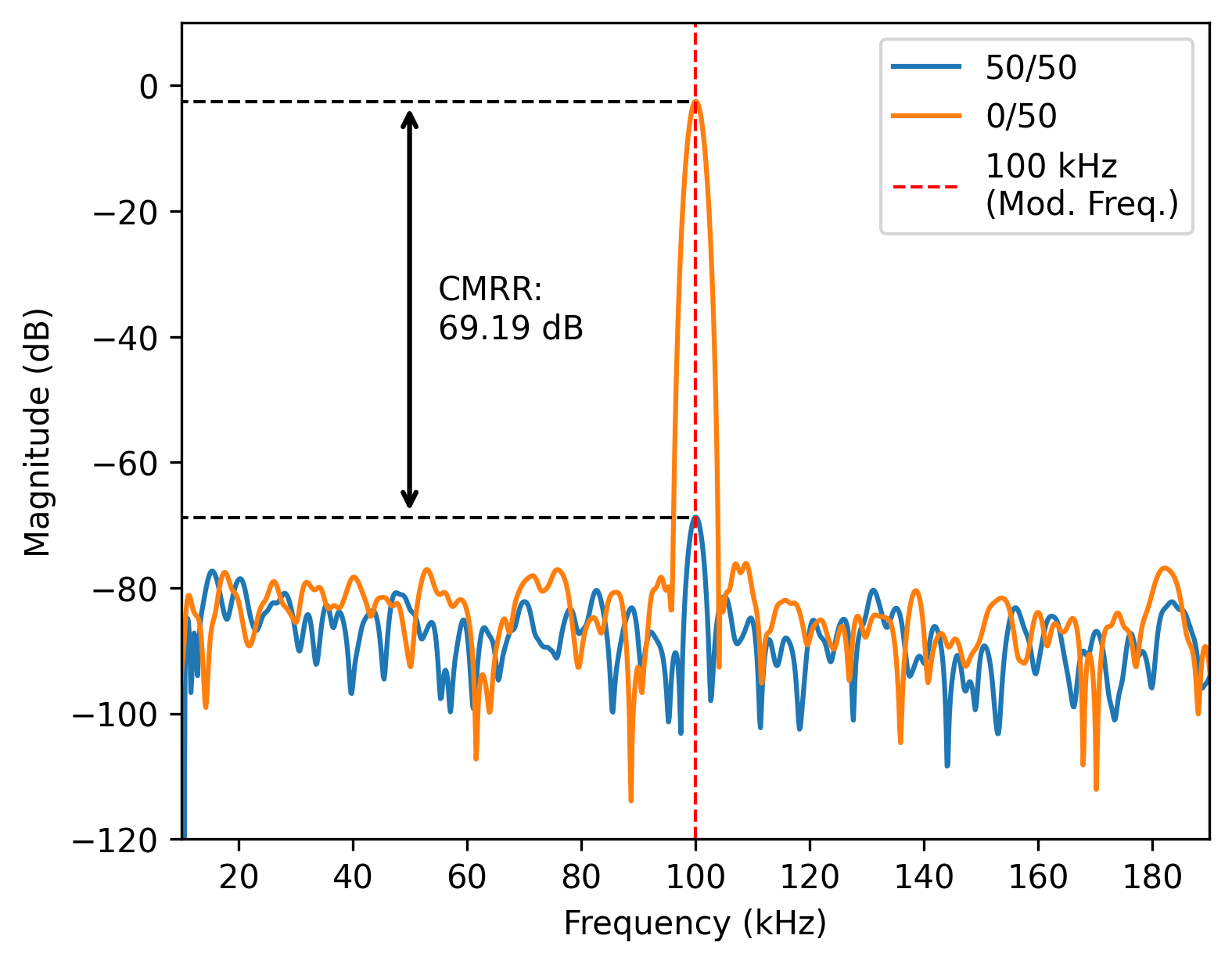}
\caption{\label{fig:cmrr}Common Mode Rejection Ratio (CMRR). A 100 kHz common mode signal is denoted by the red dashed line. The blue curve is the output in the balanced case and the orange curve is the output in the unbalanced case. The difference between the orange and blue curves at the 100 kHz modulation frequency corresponds to the optical CMRR.}
\end{figure}

To determine $S_{d}$, we fully attenuate one arm of the homodyne detector in the 0/50 configuration shown in Fig. \ref{fig:cmrr} and measure the spectral power at our modulation frequency. As this is only half of our differential signal, we need to double this power. To obtain CMRR, we take the ratio of the fully balanced 50/50 configuration to double the 0/50 configuration. This results in a CMRR of 69 dB for our system at 100 kHz. Due to limitations of thermal cross talk on the PIC between output MZMs, we consider this measurement to be a lower bound because we cannot guarantee the signal in the 0/50 configuration is fully attenuated on one arm and fully unattenuated on the other.

\subsection{Shot Noise Clearance}
\begin{figure}[t!]
\includegraphics[width=0.45\textwidth]{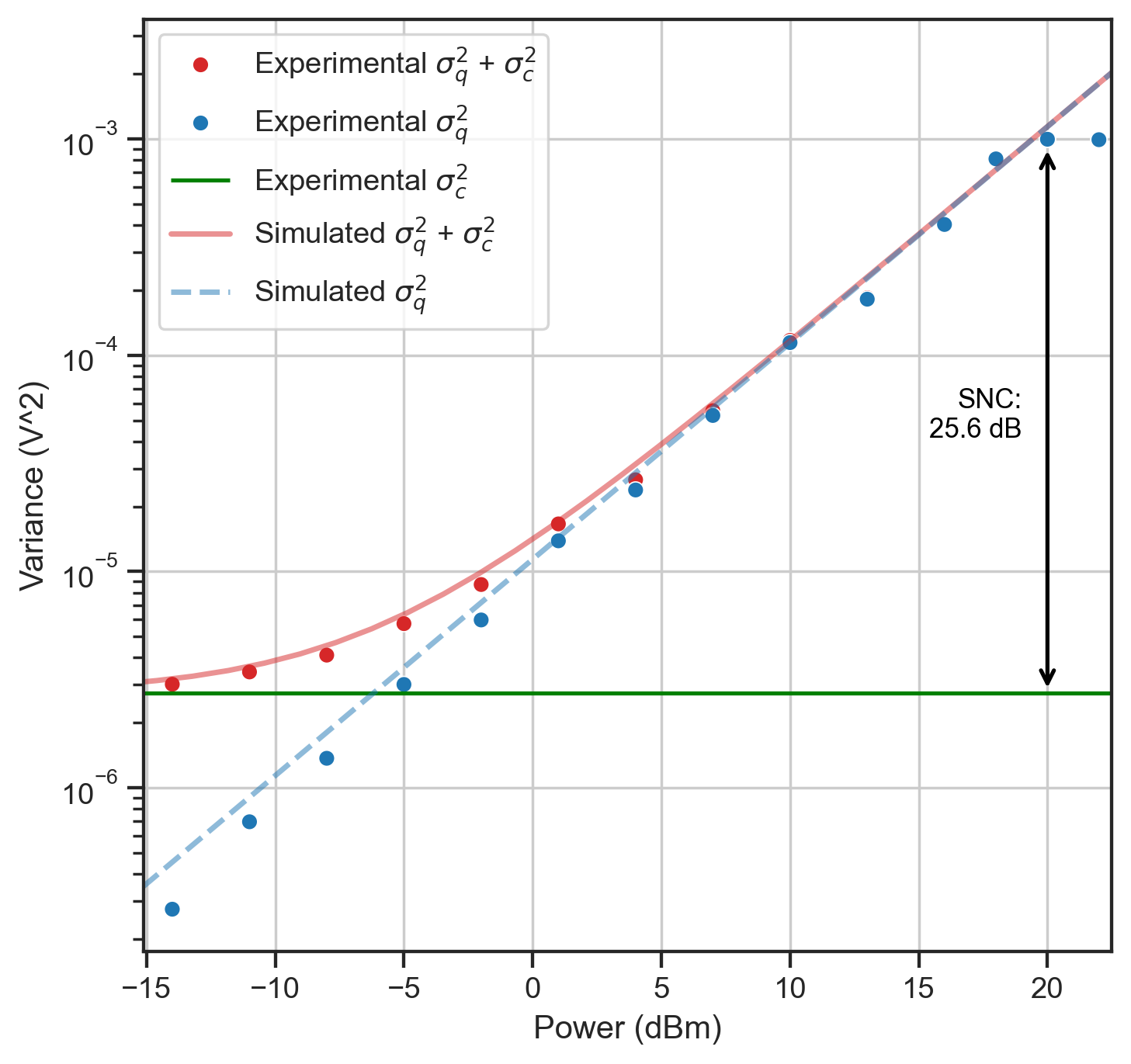}
\caption{\label{fig:snc}Shot Noise Clearance. The green line is the variance of the classical noise and the red data points represent the measured system variance at increasing LO power levels. The blue data points show the variance of the quantum noise points, and are obtained by subtracting the classical noise floor from the system variance. The red and blue lines provide a comparison to simulated total and quantum noise variance, respectively.}
\end{figure}

A common performance metric for homodyne detection is a system's shot noise clearance (SNC) \cite{integrated-homodyne-best, integrated-homodyne}, defined by
\begin{equation}
    \text{SNC} = \frac{\sigma_q^2}{\sigma_c^2},
    \label{SNC}
\end{equation}
where $\sigma_q^2$ is the variance of the quantum shot noise and $\sigma_c^2$ is the variance of the classical noise or electrical noise floor of the detector. This is equivalent to the system's SNR, referenced in the introduction. First, we take a measurement of the system with no LO power to obtain the noise floor, $\sigma_c^2$, of the system, shown as a green line in Fig. \ref{fig:snc}. We then conduct an SNC measurement with the vacuum state at multiple LO power levels ranging from -14 dBm to 22 dBm. The red data points in Fig. \ref{fig:snc} correspond to the raw detector output variance and the blue data points correspond to $\sigma_q^2$, or the difference between the raw detector variance and the detector noise floor. 

We compare our measured results to our PIC simulation results, which show good consensus. We also note that near the maximum power of our LO range we see drop off in shot noise variance and a deviation from simulation results. This is due to photodiode saturation limits not modeled in our PIC simulation tool. We report a shot noise clearance of 25.6 dB just before our photodiodes reach saturation at an LO power of 20 dBm.

\section{Randomness Extraction and Post Processing}
\subsection{Toeplitz Hashing Extractor}

\begin{figure*}[t!]
\includegraphics[width=\linewidth]{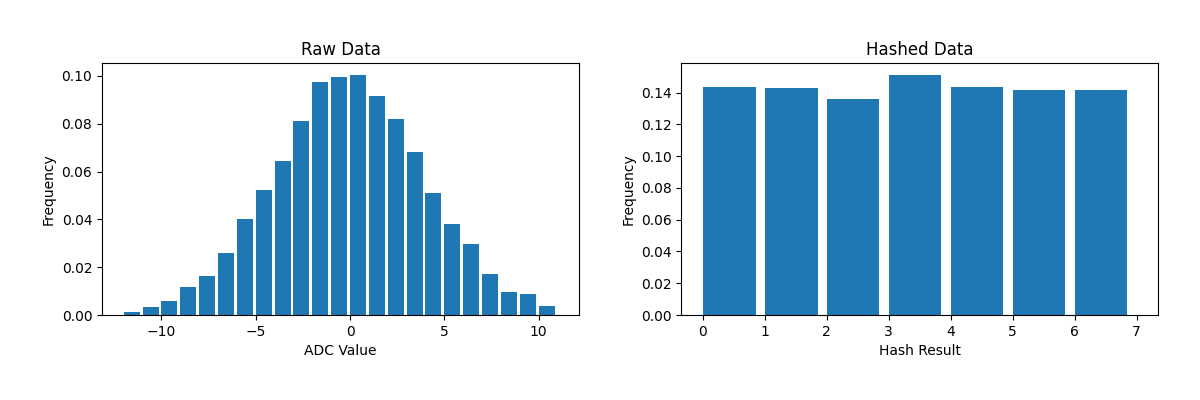}
\caption{\label{fig:hash}Toeplitz Hash Example. The left distribution shows random data from the ADC, while the right shows random numbers generated after Toeplitz hashing.}
\end{figure*}

After extracting the random samples from the TIA and measuring with an ADC, we extract quantum random numbers from the resulting distribution. In this paper, we perform QRNG with both Gaussian ($\hat{\text{X}}$ and $\hat{\text{P}}$) and chi-squared ($\hat{\text{Z}}$) distributions. Due to electronic noise in the system, these distributions contain a mix of classical and quantum noise. We implement real-time randomness extraction to extract only the contribution of the quantum noise and generate a uniform distribution. This is illustrated in Fig. \ref{fig:hash}.

In order to determine how many bits can be extracted from our input signal, we find the min-entropy, which is a conservative underestimate of the signal's entropy. The following is a derivation of the min-entropy of our system. Our method is similar to those performed in papers by Ma, et al. \cite{ma2013postprocessing} and Xu, et al \cite{xu2012ultrafast}. Here we provide a more thorough explanation of the derivation.

For a random variable $X$ composed of n random bits ($X \in \{0,1\}^n$), the general equation for min-entropy $H_{min}$ is given by

\begin{equation}
H_{min}(X) = -\mathrm{log_2}\left(\underset{v\in\left\{0,1\right\}^n}{\mathrm{max}}\;\mathrm{Prob}[X=v]\right)
\label{min-entropy}
\end{equation}.

It is important to calculate only the min-entropy of the quantum signal, $\sigma_{q}^2$. This variance can be found in Fig. \ref{fig:snc}. Next, we find the maximum of the signal's distribution. For a Gaussian, this is
\begin{equation}
\max{\mathcal{N}(\mu, \sigma_{q}^2)} = \frac{1}{\sigma_{q}\sqrt{2\pi}}.
\label{max_gaussian}
\end{equation}

To convert this to a probability, we need the ADC bin width $W_{bin}$, which is found by consulting the ADC data sheet. Because $W_{bin}$ is so small compared to $\sigma_q$ due to the high gain of our system, we can use these parameters to make a near perfect approximation for the maximum probability of our ADC measurement distribution. Using this, the min-entropy is determined to be

\begin{equation}
H_{min}(\mathcal{N}) = -\mathrm{log_2}\left(\frac{W_{bin}}{\sigma_{q}\sqrt{2\pi}}\right).
\label{min-entropy_deriv}
\end{equation}

This gives the min-entropy of our $\hat{\text{X}}$ and $\hat{\text{P}}$ measurements. This is also a suitable lower bound for the $\hat{\text{Z}}$ samples. Once the min-entropy is found, an appropriate hash family can perform randomness extraction on our input signal. Similar to other QRNG implementations, such as those by Raffaelli, et al. \cite{integrated-homodyne} and Xu, et al. \cite{xu2012ultrafast}, we implement the Toeplitz hashing extractor. Toeplitz hashes are a two-universal hash family from $n$ to $m$ bits ($m < n$), generated from an $m + n - 1$ bit random seed. In our implementation, $n$ corresponds to a single sample size and $m$ corresponds to the number of bits extracted based on the min-entropy.

Unlike other implementations, we characterize the uniformity of our Toeplitz hash using a method similar to that of Ma, et al. \cite{ma2013postprocessing}. By the Leftover Hash Lemma \cite{10.1145/73007.73009}, a distribution generated from a random $m \times n$ hash taken from a two-universal hash family and a sample with min-entropy $H_{min}(X)=k$ is statistically close to a uniform distribution, with a statistical difference bounded by $2^{-(k-m)/2}$. Therefore, the error decreases when the signal's min-entropy, $k$, is much greater than the extracted bits, $m$.

In order to increase this difference, a common technique is to concatenate samples \cite{7543094, xu2012ultrafast}. Consider a new random variable $X’$ which is formed by concatenating $s$ samples of $X$. This new random variable takes values $v’$ in the set $\{0,1\}^{s*n}$. If the original variable $X$ has entropy $H(X)$, $X’$ has entropy $s * H(X)$. If we use a $sm \times sn$ Toeplitz matrix to extract bits from this concatenated distribution, the error $\epsilon_{toeplitz}$ is bounded by

\begin{equation}\label{error}
\epsilon_{toeplitz} \leq {2^{s*(m-H_{min}(X))/2}}.
\end{equation}
Therefore, the number of bits that can be extracted from each sample while maintaining a low error increases with the number of samples $s$.

\subsection{FPGA Implementation}
The Toeplitz hashing extractor described in the previous section is implemented on an FPGA for real-time randomness extraction using methods similar to work done by Zhang, et al. \cite{7543094}. We use a Basys3 development board with a Xilinx Artix-7 FPGA core. Appendix \ref{sec:appendix_b} can be consulted for a more complete description of the FPGA architecture, but the key points are highlighted here. 

The Basys3 development board utilizes a 1 MSPS, 12-bit XADC configured in bipolar mode for a voltage range of -500 mV to 500 mV. This ADC samples the TIA output for either the X or P quadrature for use in QRNG. From there, the result is hashed in real time and send to a computer via an FT2232 USB chip.

From Fig. \ref{fig:snc}, the maximum total variance and SNC of the TIA output occur at 20 mW and are equal to 995 $\mathrm{mV^2}$ and 25.6 dB, respectively. We can use Eq. \ref{SNC} to determine the variance of just the quantum contribution. Evaluating Eq. \ref{min-entropy_deriv} with these parameters yields a min-entropy of 8.312. This gives an upper limit on the number of uniform random bits that can be extracted from each sample. The FPGA implementation extracts 8 bits from each sample and concatenates 60 consecutive samples for each hash in order to maximize uniformity. With an ADC sampling rate of 1 MSPS, this gives us a real-time generation rate of quantum random numbers of 8 Mbps. Using Eq. \ref{error}, we find $\epsilon_{toeplitz} \leq 0.0015$.

For testing purposes, data can be sent over the FPGA's USB channel to be hashed and returned. Additionally, the raw ADC signal can be returned over USB. Using these features, individual sections of the FPGA implementation were tested to confirm accuracy and reliability. All of the FPGA logic is synchronized with the Basys3 100 MHz clock signal, and the implementation is capable of hashing one sample per clock cycle.

\subsection{Python Implementation}
The FPGA implementation of the Toeplitz hashing extractor was used to collect, hash, and output the $\hat{\text{X}}$ and $\hat{\text{P}}$ data of the system. In order to verify the results of the FPGA implementation, the algorithm outlined in the previous two sections was also implemented in Python code. The FPGA was used to collect and store raw $\hat{\text{X}}$ and $\hat{\text{P}}$ samples as binary files, which were used as inputs to the Python implementation. Additionally, we squared these measurements and added them together to produce a measurement of the $\hat{\text{Z}}$ operator, and Toeplitz hashing was performed on the result.

\subsection{NIST and Bitrate}
In order to authenticate the genuine randomness of the hashed $\hat{\text{X}}$, $\hat{\text{P}}$, and $\hat{\text{Z}}$ distributions, we ran our data through the National Institue of Standards and Technology (NIST) Statistical Test Suite for Random and Pseudorandom Number Generators \cite{rukhin2001statistical}. 10 MB samples of the hashed $\hat{\text{X}}$ and $\hat{\text{P}}$ distributions passed all tests in the NIST test suite. $\hat{\text{X}}$ and $\hat{\text{P}}$ samples of approximately 4.8 MB were then used as inputs to the Python implementation of the Toeplitz hashing extractor to produce a $\hat{\text{Z}}$ distribution, which also passed all tests in the suite. Using Fisher's method, the p-values of the $\hat{\text{X}}$, $\hat{\text{P}}$, and $\hat{\text{Z}}$ distributions for each test were condensed into one composite p-value, shown in Table \ref{table:fisher}. These p-values were compared against a significance level, $\alpha$, of 0.01. A more complete summary of these tests and the resulting p-values of the $\hat{\text{X}}$, $\hat{\text{P}}$, and $\hat{\text{Z}}$ distributions for each individual test can be found in Appendix \ref{sec:appendix_c}. 

The $\hat{\text{Z}}$ distribution has a higher p-value because the samples have a much higher entropy, but 8 bits were still extracted from each sample. Though the significance values for each test were only slightly higher, applying Fisher's method amplified this difference. All three of these distributions have p-values above the significance level of 0.01. We can therefore demonstrate the genuine randomness of our QRNG. Given that our system uses a 1 MSPS ADC and 1 byte is extracted from each sample, the bitrate of our QRNG is 8 Mbps, and is primarily limited by the ADC sampling speed and not the TIA bandwidth. Because our system has a maximum bandwidth of 100 MHz, with an improved post-processing setup, the theoretical limit on our bitrate is 800 Mbps.

\renewcommand{\arraystretch}{1.5}
\begin{table}[h!]
\centering
\begin{tabular}{|>{\raggedright\arraybackslash}p{3cm}|p{3cm}|}
\hline
Distribution & P-Value \\
\hline
$\hat{\text{X}}$ & 0.293603109480466 \\
\rowcolor{lightgray!25}$\hat{\text{P}}$ & 0.886935464546968 \\
$\hat{\text{Z}}$ & 0.999963450017898 \\
\hline
\end{tabular}
\caption{Composite P-values for $\hat{\text{X}}$, $\hat{\text{P}}$, and $\hat{\text{Z}}$.}
\label{table:fisher}
\end{table}

\section{Conclusion}
In this work we have demonstrated a novel integrated differential homodyne detection scheme for QRNG. Our system achieved an SNC of 25.6 dB, a CMRR of 69 dB, a QRNG real time bitrate of 8 Mbps, and a theoretical bitrate of 800 Mbps. Additionally, we proposed using $\hat{\text{Z}}$ as a source of randomness. This approach has a small advantage in min-entropy compared to a single vacuum homodyne detection scheme, and has the additional benefit of being phase independent. We chose vacuum state homodyne detection QRNG to demonstrate the capabilities of this detector, but there are other applications in which this type of device would exceed the performance of traditional methods. \pagebreak For example, by utilizing $\hat{\text{Z}}$, one can perform QRNG on super-Poissonian states to achieve even higher levels of min-entropy and SNC. Our system could also be very useful in other quantum information applications such as QKD.

\bibliographystyle{IEEEtran}
\bibliography{references}

\clearpage

\appendix

\section{\label{sec:appendix_a}Derivations}
\subsection{CMRR Due to Path Length Difference}
The following is a derivation for optical common mode rejection ratio (CMRR) due to path length difference.

The general equation for CMRR is given by
\begin{equation}
\text{CMRR} = 10\log_{10}\left(\frac{S_{d}}{S_{cm}}\right).
\label{CMRR}
\end{equation}
Where $S_d$ is the differential signal and $S_{cm}$ is the common mode signal. To simplify the derivation, this can be normalized by setting $S_{d}$ equal to 1 and moving $S_{cm}$ to the numerator of the logarithmic.

\begin{equation}
\text{CMRR} = -10\log_{10}\left(S_{cm}\right)
\label{CMRR_norm}
\end{equation}

We consider a signal that is evenly split into two paths. The result is two waves of equal frequencies, each propagating down a separate branch. If the path lengths are imbalanced, a phase difference is accumulated between the two waves. This results in an amplitude difference between the two signals, denoted $S_{cm}$.

To find the amplitude difference between two sinusoids of equal frequency and differing phase, we must first define the wave in each branch. We will designate $x_{1}(t)$ as a sinusoid with a phase of $\phi_{1}$ and $x_{2}(t)$ as a sinusoid with a phase of $\phi_{2}$.
\begin{equation}
x_{1}(t) = A_{1}\sin{\left(\omega t + \phi_{1}\right)}
\label{sine_1}
\end{equation}
\begin{equation}
x_{2}(t) = A_{2}\sin{\left(\omega t + \phi_{2}\right)}
\label{sine_2}
\end{equation}
We then calculate the difference between Eq. \ref{sine_1} and Eq. \ref{sine_2}, denoted $x(t)$.
\begin{equation}
x(t) = x_{1}(t) - x_{2}(t)
\label{diff_x}
\end{equation}
\begin{equation}
x(t) = A_{1}\sin{\left(\omega t + \phi_{1}\right)} - A_{2}\sin{\left(\omega t + \phi_{2}\right)}
\label{diff}
\end{equation}
Assuming an ideal optical splitting ratio, $A_{1}$ is equal to $A_{2}$. This value can be factored out and normalized to 0.5, leaving
\begin{equation}
x(t) = 0.5\left(\sin{\left(\omega t + \phi_{1}\right)} - \sin{\left(\omega t + \phi_{2}\right)}\right).
\label{diff_factored}
\end{equation}
Using the sum-to-product trigonometric identity for the difference of sines, defined as
\begin{equation}
\resizebox{0.42\textwidth}{!}{$\sin{\left(\alpha\right)} - \sin{\left(\beta\right)} = 2\cos{\left(\frac{\left(\alpha + \beta\right)}{2}\right)}\sin{\left(\frac{\left(\alpha - \beta\right)}{2}\right)},$}
\label{trig_ID}
\end{equation}
$x(t)$ can be equivalently represented as
\begin{equation}
\resizebox{0.42\textwidth}{!}{$x(t) = \cos{\left(\frac{\left(\omega t + \phi_{1}\right)+\left(\omega t + \phi_{2}\right)}{2}\right)}\sin{\left(\frac{\left(\omega t + \phi_{1}\right)-\left(\omega t + \phi_{2}\right)}{2}\right)}.$}
\label{trig_ID_sub}
\end{equation}
This simplifies to
\begin{equation}
x(t) = \cos{\left(\omega t + \frac{\phi_{1}+\phi_{2}}{2}\right)}\sin{\left(\frac{\phi_{1}-\phi_{2}}{2}\right)}
\label{simplify}
\end{equation}
or, equivalently,
\begin{equation}
x(t) = \sin{\left(\frac{\phi_{1}-\phi_{2}}{2}\right)}\cos{\left(\omega t + \frac{\phi_{1}+\phi_{2}}{2}\right)}.
\label{swap}
\end{equation}
The amplitude of $x(t)$, which we previously defined to be $S_{cm}$, can then be easily identified as the magnitude of the coefficient of the oscillating cosine.
\begin{equation}
S_{cm} = \abs{\sin{\left(\frac{\phi_{1}-\phi_{2}}{2}\right)}}
\label{amplitude}
\end{equation}
and the phase difference can be redefined as $\Delta\phi$.

\begin{equation}
S_{cm} = \abs{\sin{\left(\frac{\Delta\phi}{2}\right)}}
\label{amplitude_delta}
\end{equation}

We now have an equation for the difference in the amplitude of the two waves, given their phase difference. The next step is to derive $\Delta\phi$. Given a signal with a modulation frequency $f_{cm}$ in a medium of length $L$ with refractive index $n_{eff}$, the effective wavelength can be denoted as 
\begin{equation}
\lambda_{eff} = \frac{c}{fn_{eff}},
\label{lambda_eff}
\end{equation}
where $c$ is the speed of light. The accumulated phase of the wave in this medium can then be defined using its $\lambda_{eff}$.

\begin{equation}
\phi = \frac{2\pi L}{\lambda_{eff}}
\label{lambda_eff}
\end{equation}
Substituting $f_{cm}$ back into the equation yields
\begin{equation}
\phi = \frac{2\pi f_{cm}}{c}n_{eff}L.
\label{phi}
\end{equation}
Therefore, the phase difference given a path length difference $\Delta L$ can similarly be defined with a simple adjustment to Eq. \ref{phi}.

\begin{equation}
\Delta\phi = \frac{2\pi f_{cm}}{c}n_{eff}\Delta L
\label{delta_phi}
\end{equation}

Thus, the CMRR due to path length difference is 
\begin{equation}
\text{CMRR} = -10\log_{10}\left(\abs{\sin{\left(\frac{\Delta\phi}{2}\right)}}\right),
\label{CMRR_final}
\end{equation}
where $\Delta\phi$ is given by Eq. \ref{delta_phi}.

\begin{figure*}[t!]
\includegraphics[width=\linewidth]{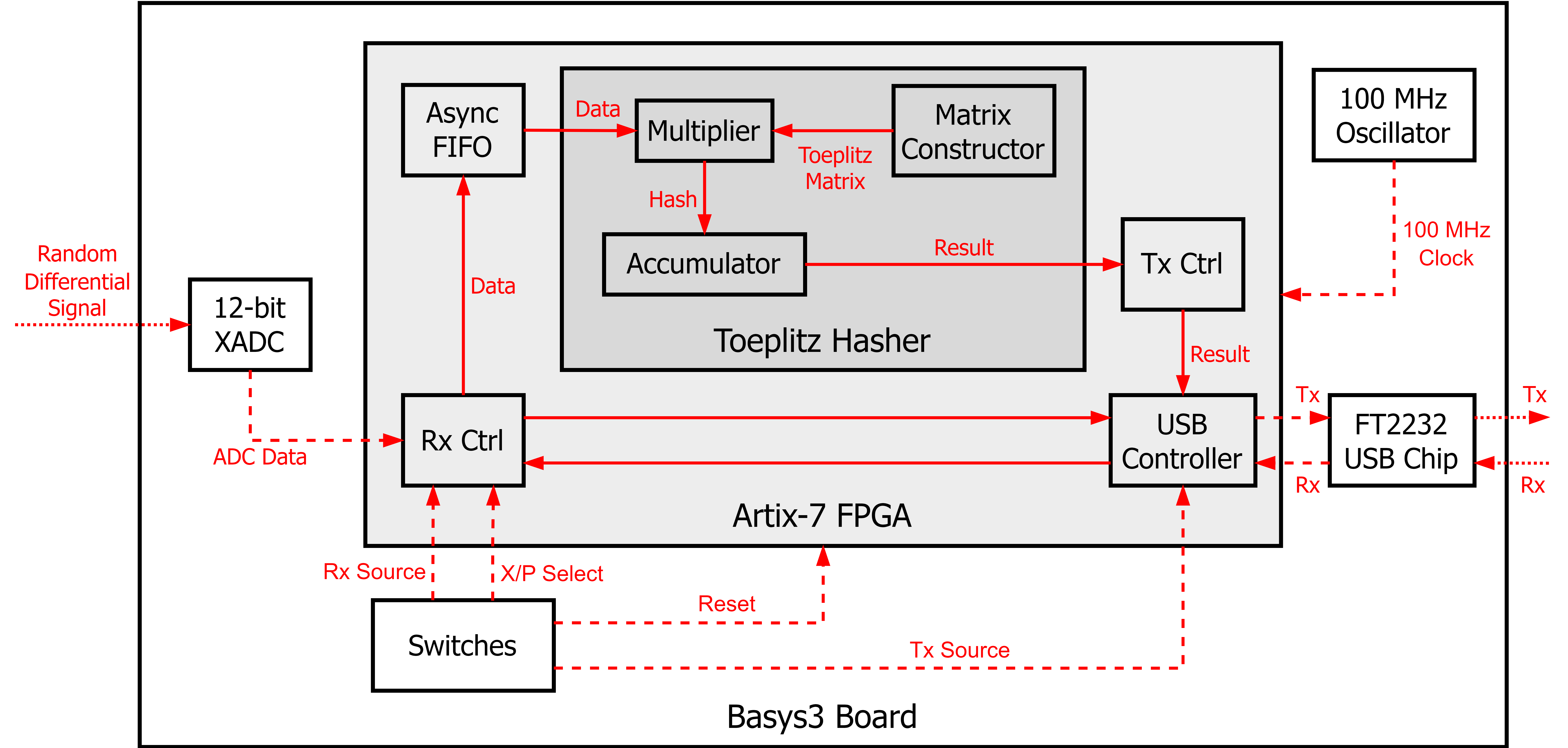}
\caption{\label{fig:fpga}FPGA Block Diagram.}
\end{figure*}

\subsection{CMRR Due to Optical Splitting Imbalance}
The following is a derivation for optical CMRR due to optical splitting imbalance. 

We begin this derivation with the same normalized equation for CMRR given by Eq. \ref{CMRR_norm}. We then define the power coupling coefficient, $\kappa$, as
\begin{equation}
\kappa = \frac{x_{1}}{x_{1}+x_{2}},
\label{kappa}
\end{equation}
where $x_{1}$ and $x_{2}$ are inputs to ports 1 and 2 of an amplifier, respectively. A $\kappa$ value of 0.5 indicates an equal splitting ratio and maximum CMRR, while a $\kappa$ value of 0 or 1 indicates the worst case, in which all optical power is concentrated on one port and CMRR is minimized. We relate this value to $S_{cm}$ using the transmission coefficient, $T$, defined as
\begin{equation}
T = \frac{x_{2}}{x_{1}+x_{2}}.
\label{transmission}
\end{equation}
Because $\kappa$ and T sum to 1, the positive difference between these two values quantifies the optical splitting imbalance and can be denoted as $S_{cm}$.
\begin{equation}
S_{cm} = \abs{\kappa - T}
\label{difference}
\end{equation}
In terms of $x_{1}$ and $x_{2}$, this can be expanded to
\begin{equation}
S_{cm} = \abs{\frac{x_{1}}{x_{1}+x_{2}} - \frac{x_{2}}{x_{1}+x_{2}}} = \frac{\abs{x_{1}-x_{2}}}{x_{1}+x_{2}}.
\label{difference_x}
\end{equation} 
This can also be expressed in terms of only $\kappa$ as
\begin{equation}
S_{cm} = \abs{\kappa - \left(1 - \kappa\right)} = \abs{2\kappa - 1}.
\label{difference_kappa}
\end{equation}

Therefore, the CMRR due to optical splitting imbalance is
\begin{equation}
\text{CMRR} = -10\log_{10}\left(\frac{\abs{x_{1}-x_{2}}}{x_{1}+x_{2}}\right)
\label{CMRR_final_x}
\end{equation}
or, equivalently,
\begin{equation}
\text{CMRR} = -10\log_{10}\left(\abs{2\kappa - 1}\right).
\label{CMRR_final_kappa}
\end{equation}

\section{\label{sec:appendix_b}FPGA Architecture}
Fig. \ref{fig:fpga} shows a block diagram of our FPGA configuration. In this appendix, a brief overview of each part of this block diagram is given.

After leaving the TIA, random signals are sampled using a 12-bit XADC. Using the X/P Select switch, the ADC performs either $\hat{\text{X}}$ or $\hat{\text{P}}$ measurements. The digitized signal is sent to the Rx Ctrl module. Using the Rx Source switch, this module selects either the ADC measurements or bit samples received from the USB Controller for testing purposes. These selected values are sent to an asynchronous FIFO as they arrive. 

\begin{figure}[t!]
\includegraphics[width=0.45\textwidth]{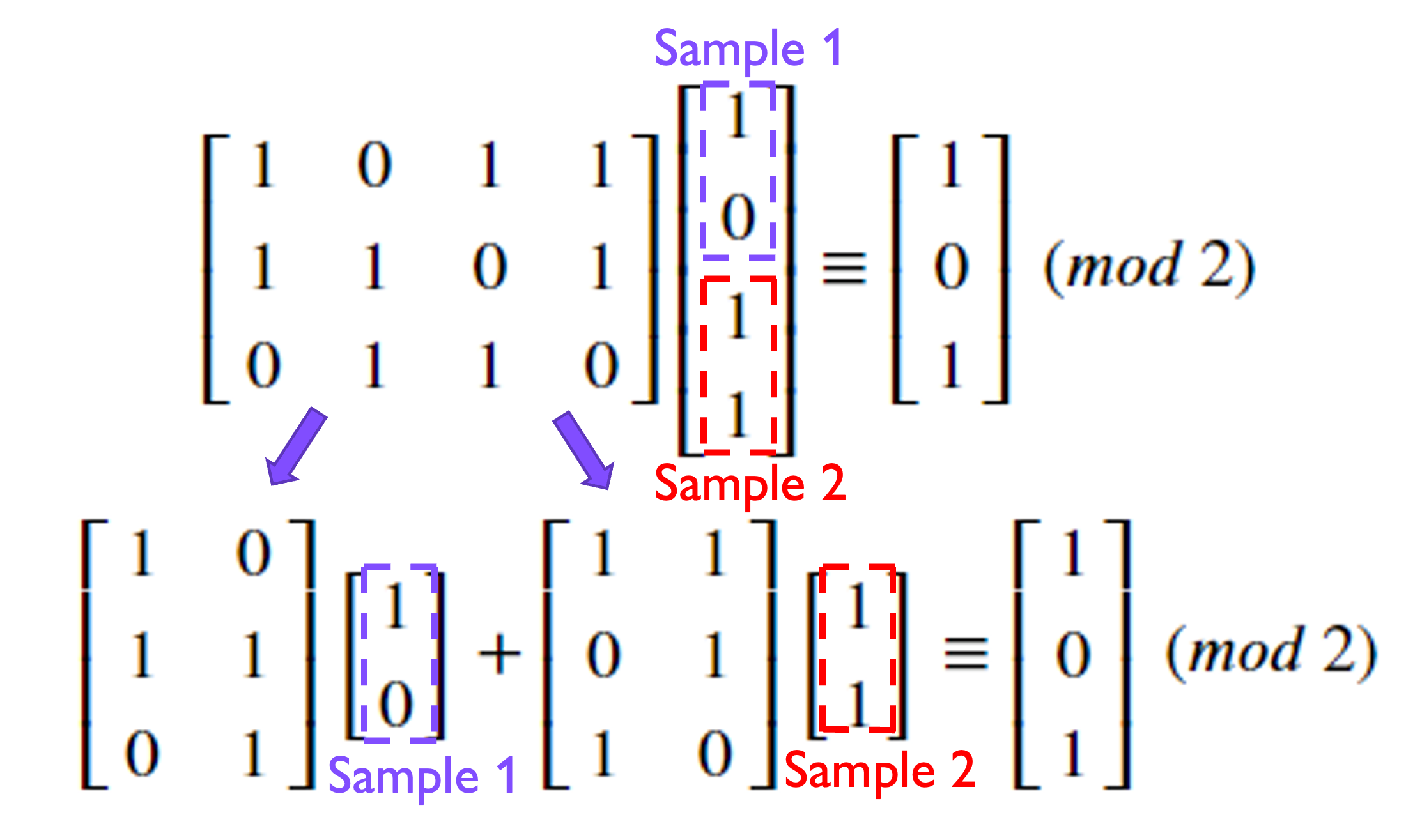}
\caption{\label{fig:slices}Hashing with Slices Example. In this example, an $(sn) \times (sm)$ Toeplitz matrix is used to compute the hash as samples arrive, where $s = 2$, $m = 2$, and $n = 1.5$.}
\end{figure}

The FPGA uses an $(sn) \times (sm)$ Toeplitz matrix to generate uniformly distributed random numbers, where $s$ is the number of samples to concatenate for one hash. This matrix is constructed from a random seed of size $s*(m + n) - 1$ bits. In order to do this operation in real time, the full Toeplitz matrix is divided into slices for each sample as shown in Fig. \ref{fig:slices}. As samples arrive, they are multiplied by their associated $(sn) \times m$ slice of the full Toeplitz matrix. Matrix multiplication can then be performed using a bitwise AND operation and results from consecutive samples can be added using a bitwise XOR operation.

Once every clock cycle, the Toeplitz Hasher module checks for available data on the asynchronous FIFO. If data is available, it is immediately multiplied by its associated slice of the full Toeplitz matrix, which is generated by the Matrix Constructor module. The Accumulator module collects each result and sends the final hash to Tx Ctrl after $s$ samples have arrived.

The Tx Ctrl module and the USB Controller module work together to send data off the board. Using a switch on the Basys3 board (labeled Tx Source), the USB Controller trasmits either the resulting hash from the Toeplitz Hasher module or the raw data from Rx Ctrl for testing. The FT2232 USB Chip on the Basys3 board is used to send information to a computer. Additionally, the USB Controller can receive data from the computer to send to Rx Ctrl for testing.

With different combinations of Rx Source and Tx Source, the board has four modes of operation: 
\begin{enumerate}
  \item Hash and transmit ADC data (Normal Operation)
  \item Receive, hash, and transmit USB data (Hash Testing)
  \item Transmit ADC data (ADC Testing)
  \item Receive and transmit USB data (USB Testing)
\end{enumerate}

When compared against our Python implementation of Toeplitz hashing, this allows us to confirm reliability and accuracy for every piece of the system.

\section{\label{sec:appendix_c}NIST Test Results}
Table \ref{table:NIST} presents a summary of the NIST randomness tests passed by our QRNG system, along with the resulting p-values of the $\hat{\text{X}}$, $\hat{\text{P}}$, and $\hat{\text{Z}}$ distributions for each test, compared against a significance level, $\alpha$, of 0.01. The $\hat{\text{X}}$ and $\hat{\text{P}}$ samples used in the test suite were each 10 MB. Additional 4.8 MB $\hat{\text{X}}$ and $\hat{\text{P}}$ samples were collected to produce the $\hat{\text{Z}}$ distribution used in the test suite.

\setlength{\arrayrulewidth}{0.5mm}
\setlength{\tabcolsep}{8pt}
\renewcommand{\arraystretch}{1.45}

\begin{table*}[htbp]
\centering
\scriptsize
\begin{tabular}{|>{\raggedright\arraybackslash}p{3cm}|p{3cm}|p{3cm}|p{3cm}|}
\hline
Test & $\hat{\text{X}}$ P-Value & $\hat{\text{P}}$ P-Value & $\hat{\text{Z}}$ P-Value \\
\hline
Frequency Test (Monobit) & 0.14250526321366025 & 0.858378832079647 & 0.5326402202519169 \\
\rowcolor{lightgray!25}Frequency Test within a Block & 0.05325581878678178 & 0.4958594427147481 & 0.6981191634047049 \\
Run Test & 0.09855246668255926 & 0.581960365722031 & 0.2208476576515993 \\
\rowcolor{lightgray!25}Longest Run of Ones in a Block & 0.015828516441345104 & 0.6966116483914697 & 0.9401805373576853 \\
Binary Matrix Rank Test & 0.1031406805234125 & 0.7558957960420669 & 0.7435322031345005 \\
\rowcolor{lightgray!25}Discrete Fourier Transform (Spectral) Test & 0.06411151194654537 & 0.026126850266324296 & 0.9986410717331123 \\
Non-Overlapping Template Matching Test & 0.07324177424496794 & 0.5769493230931108 & 0.11599617467068 \\
\rowcolor{lightgray!25}Overlapping Template Matching Test & 0.04443336097391225 & 0.05441057035109533 & 0.1477881048240294 \\
Maurer's Universal Statistical Test & 0.028655128056009168 & 0.6768092549346105 & 0.5454111913405972 \\
\rowcolor{lightgray!25}Linear Complexity Test & 0.9124435615958901 & 0.7553593546698116 & 0.9077251265662214 \\
\multirow{2}{3cm}{Serial Test} & 0.4453466441754355 & 0.6534911169303284 & 0.3702524403470766 \\
 & 0.43600089278880055 & 0.1654228490860131 & 0.648414319419013 \\
\rowcolor{lightgray!25}Approximate Entropy Test & 0.5110510326615959 & 0.9578657025360328 & 0.34647475241710685 \\
Cumulative Sums (Forward) Test & 0.0837601068767868 & 0.5838230301860956 & 0.5967960412274258 \\
\rowcolor{lightgray!25}Cumulative Sums (Reverse) Test & 0.0837601068767868 & 0.5838230301860956 & 0.5967960412274258 \\
\hline
\multicolumn{4}{|l|}{Random Excursions Test} \\
\hline
State & $\hat{\text{X}}$ P-Value & $\hat{\text{P}}$ P-Value & $\hat{\text{Z}}$ P-Value \\
\hline
-4 & 0.8922082121681362 & 0.2576505914389607 & 0.23112295574423738 \\
\rowcolor{lightgray!25}-3 & 0.9956139024768987 & 0.5162769876977834 & 0.8884263033334616 \\
-2 & 0.4201628510333726 & 0.9926733389355058 & 0.9533269458013858 \\
\rowcolor{lightgray!25}-1 & 0.7720843766425737 & 0.8507119188007102 & 0.7480446895783766 \\
+1 & 0.9844363651859271 & 0.6926918134797209 & 0.43382713849087673 \\
\rowcolor{lightgray!25}+2 & 0.9367806423198609 & 0.01724247612960804 & 0.45498355128670787 \\
+3 & 0.9148935431673655 & 0.17477136963206968 & 0.7860218155711133 \\
\rowcolor{lightgray!25}+4 & 0.6945333185229909 & 0.4767343072185293 & 0.9997036284825281 \\
\hline
\multicolumn{4}{|l|}{Random Excursions Variant Test} \\
\hline
State & $\hat{\text{X}}$ P-Value & $\hat{\text{P}}$ P-Value & $\hat{\text{Z}}$ P-Value \\
\hline
-9 & 0.43191782847429594 & 0.15169113484743674 & 0.9091147574701322 \\
\rowcolor{lightgray!25}-8 & 0.5599653900967534 & 0.2674994624172822 & 0.8125583981807282 \\
-7 & 0.9609330364549469 & 0.4254836262322358 & 0.7047705345308095 \\
\rowcolor{lightgray!25}-6 & 0.8730783144590134 & 0.37986739103989997 & 0.789838325196611 \\
-5 & 0.8477678793532261 & 0.31007362401001026 & 0.7133540050708602 \\
\rowcolor{lightgray!25}-4 & 0.7916976874221073 & 0.4359067576202531 & 0.7812468757872679 \\
-3 & 0.7520561740802294 & 0.8641412126377457 & 0.9918075506943677 \\
\rowcolor{lightgray!25}-2 & 0.18943418792053068 & 0.9892095626844404 & 0.7403462770162056 \\
-1 & 0.1363121685834667 & 0.7607315539146968 & 0.6710155965874569 \\
\rowcolor{lightgray!25}+1 & 0.7181754138113141 & 0.8208606605533411 & 0.5128852839129775 \\
+2 & 0.8836856129346609 & 0.6651782104904482 & 0.6666041527699841 \\
\rowcolor{lightgray!25}+3 & 0.8501940991342305 & 0.3676307772791674 & 0.758054339185876 \\
+4 & 0.9236993020300784 & 0.460631931447131 & 0.814748404582641 \\
\rowcolor{lightgray!25}+5 & 0.8698791201341265 & 0.8759035632325761 & 0.9055723384819724 \\
+6 & 0.6284715509496972 & 0.7757468019600118 & 0.6854975024420304 \\
\rowcolor{lightgray!25}+7 & 0.5739546398763362 & 0.9051894283396957 & 0.6397567351388128 \\
+8 & 0.5053067075488527 & 0.8041523453721473 & 0.6912292427522475 \\
\rowcolor{lightgray!25}+9 & 0.37938400377333115 & 0.5307656085817369 & 0.6864188933703024 \\
\hline
\end{tabular}
\caption{NIST Test results for $\hat{\text{X}}$, $\hat{\text{P}}$, and $\hat{\text{Z}}$.}
\label{table:NIST}
\end{table*}

\end{document}